\documentclass[journal=jpccck,manuscript=article,layout=traditional]{achemso}
\setkeys{acs}{usetitle=false}
\usepackage{natmove}
\usepackage{multirow}
\usepackage{setspace}
\usepackage{caption}
\usepackage{array}
\usepackage{booktabs}
\usepackage{rotating}
\usepackage{longtable}
\usepackage{amsmath}
\usepackage{achemso}
\setkeys{acs}{usetitle = true}
\setkeys{acs}{maxauthors=10, etalmode=truncate}
\usepackage[capitalise,noabbrev]{cleveref}
\usepackage{color}

\newif\iffigures
\figurestrue

\author{Ada Della Pia}
\affiliation{Department of Physics, La "Sapienza" University, 00185 Roma, Italy}
\email{Ada.Della.Pia@roma1.infn.it}
\author{Giulia Avvisati}
\affiliation{Department of Physics, La "Sapienza" University, 00185 Roma, Italy}
\author{Oualid Ourdjini}
\affiliation{Department of Physics, La "Sapienza" University, 00185 Roma, Italy}
\author{Claudia Cardoso}
\affiliation{Centro S3, CNR-Istituto Nanoscienze, 41125 Modena, Italy}
\author{Daniele Varsano}
\affiliation{Centro S3, CNR-Istituto Nanoscienze, 41125 Modena, Italy}
\author{Deborah Prezzi}
\affiliation{Centro S3, CNR-Istituto Nanoscienze, 41125 Modena, Italy}
\author{ Andrea Ferretti}
\affiliation{Centro S3, CNR-Istituto Nanoscienze, 41125 Modena, Italy}
\email{andrea.ferretti@nano.cnr.it}
\author{Carlo Mariani}
\affiliation{Department of Physics, La "Sapienza" University, 00185 Roma, Italy}
\author{Maria Grazia Betti}
\affiliation{Department of Physics, La "Sapienza" University, 00185 Roma, Italy}
\email{maria.grazia.betti@roma1.infn.it}

\title{Electronic Structure Evolution during the Growth of Graphene Nanoribbons on Au(110)}

\begin{document}
\tracingall

\begin{abstract}

Surface-assisted polymerization of molecular monomers into extended chains can be used as the seed of graphene nanoribbon (GNR) formation, resulting from a subsequent cyclo-dehydrogenation process. 
By means of valence-band photoemission and ab-initio density-functional theory (DFT) calculations, we investigate the evolution of molecular states from monomer 10,10'-dibromo-9,9'bianthracene (DBBA) 
precursors to polyanthryl polymers, and eventually to GNRs, as driven  by the Au(110) surface.

The molecular orbitals and the energy level alignment at the metal-organic interface are studied in depth for the DBBA precursors deposited at room temperature. 
On this basis, we can identify a spectral fingerprint of C-Au interaction in both DBBA single-layer and polymerized chains obtained upon heating.
Furthermore, DFT calculations help us evidencing that GNRs interact more strongly than DBBA and polyanthryl with the Au(110) substrate, as a result of their flatter conformation.

\end{abstract}


\newpage

\section{Introduction}
 
Molecular monomers with defined size, shape, and chemical functionality can be promising and versatile building blocks to produce nanostructured systems with atomic precision through a bottom-up approach~\cite{Barth,BarthAnnuRevPhysChem2007}. Whereas the characteristic length and shape of the nano-systems are determined by the size and geometry of the starting units, specific functions can be independently encoded in assembling the building blocks. Recently, this approach has been exploited to build up graphene nanostructures  
\cite{Muellen_review2015,Narita_Chem_Soc_Rev_2015}.
Among these, graphene nanoribbons (GNRs) are on the spotlight of the current research due to their semiconducting properties and promising applications, ranging from spintronics\cite{Son_2006,Magda2014} to nano- and opto-electronics\cite{Li_Science_2008,Bennett2013,Narita_ACSnano2014,Abbas2014,Cai2014,Chen_Nat_nano2015,Lee2015}.

One of the most used molecular precursor to achieve subnanometer-wide GNRs is 10,10'-dibromo-9,9'bianthracene (DBBA)  \cite{Muellen_review2015}. 
Depending on both the metal substrate and the deposition temperature, different intermediate species are formed, which can precede or block the GNR formation. 

In particular, the choice of surfaces more reactive than Au(111), such as copper substrates or Au(110), can be used to control the molecule-surface interaction, thereby affecting the GNR growth.
\cite{Cai_nature_2010,Simonov_JPCC_2014,Simonov_ACSnano2015,Batra_Chem_Science2014,
Denk_nat_com_2014,massimi_JPCC_2015}
For instance, the Au(110) surface, often used to obtain aligned molecular layers \cite{Fortuna_JPCC2012,Betti_SS_2004, Betti_Langmuir_2012},  gives rise to different intermediate species depending on the substrate temperature during DBBA adsorption \cite{massimi_JPCC_2015}.

Whilst the properties of the GNRs have been thoroughly 
investigated \cite{ruffieux2012,linden2012,chen2013,Denk_nat_com_2014,Batra_Chem_Science2014,Grill_Nat_Nano2012,Sode_PRB2015}, the evolution of the molecular states from the monomer precursors to the polymerized chains has received much less attention~\cite{BronnerJCP2014,Simonov_JPCC_2014,massimi_JPCC_2015,Denk_nat_com_2014}. However, a detailed knowledge of these  preliminary phases of the growth can better unveil the precursor-substrate interaction and the delicate interplay of the parameters influencing the formation of long-range ordered GNRs on metal substrates.  

In this paper, we present a combined experimental and theoretical study of the electronic structure of the DBBA molecular precursors and of the intermediate species formed after the dehalogenation and dehydrogenation processes on Au(110), following two recently-reported procedures leading to GNR formation \cite{massimi_JPCC_2015}. 

The DFT electronic structure of the isolated DBBA is compared with valence-band photoemission data for the adsorbed thin film, and taken as a reference. This allows us to spectroscopically evidence the presence of C-Au bonds at the molecular edges when a DBBA single-layer is deposited at 300 K, which is suggested as a critical feature to accomplish the polymerization procedure.
Moreover, our findings demonstrate an interaction of the GNRs (and other intermediate flat molecules) with the metal surface that is stronger than that found for both DBBA and polymer precursors. The latter indeed show molecular-like states, independent of the orientation on Au, at variance with GNRs.

\section{Results and Discussion}


\subsection{DBBA molecular states}

\begin{figure}
\centering
\iffigures
   \includegraphics[width=0.5\textwidth]{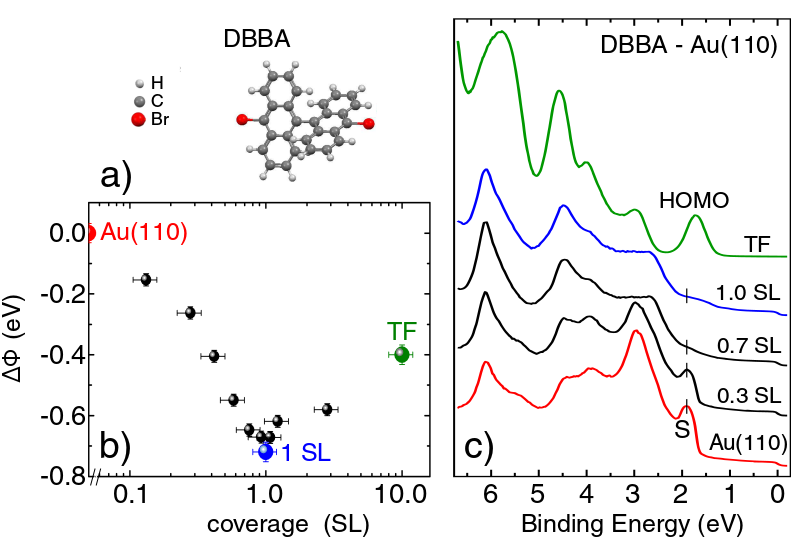}
\fi
\caption{(a) Ball-and-stick model of the DBBA molecule. (b) Work function variation ($\Delta\Phi$) with respect to the clean Au(110) substrate as a function of DBBA coverage on Au(110) at room temperature (RT) ($\Phi$ of clean Au is 5.28$\pm$0.02 eV). (c) Normal emission He-I (21.218~eV) photoemission spectra in the valence band region for the DBBA/Au(110) system, at increasing molecular coverage. Spectra are vertically displaced for clarity. S indicates the position of the Au surface state, HOMO is the highest-occupied molecular-orbital, SL and TF stand for single layer and thin film, respectively.  
\label{EDC_coverage_dependence_RT}}
\end{figure}

The evolution of the valence electronic states of DBBA deposited on Au(110) was inspected by photoemission spectroscopy at increasing molecular density, from sub-monolayer coverage up to the formation of a thin film (TF) of about 10 layers (see Methods), as shown in Fig.~\ref{EDC_coverage_dependence_RT}. 
As reported in Fig.~\ref{EDC_coverage_dependence_RT}b, the work function ($\phi$) decreases for increasing DBBA coverage, until it reaches a minimum value of 4.57 eV at the completion of the single-layer (SL). By further increasing the coverage, the work function increases again and finally reaches a saturation value of 4.87 eV, corresponding to the work function of a DBBA-TF on Au(110). 

\begin{figure}
\centering
\iffigures
   \includegraphics[width=0.5\textwidth]{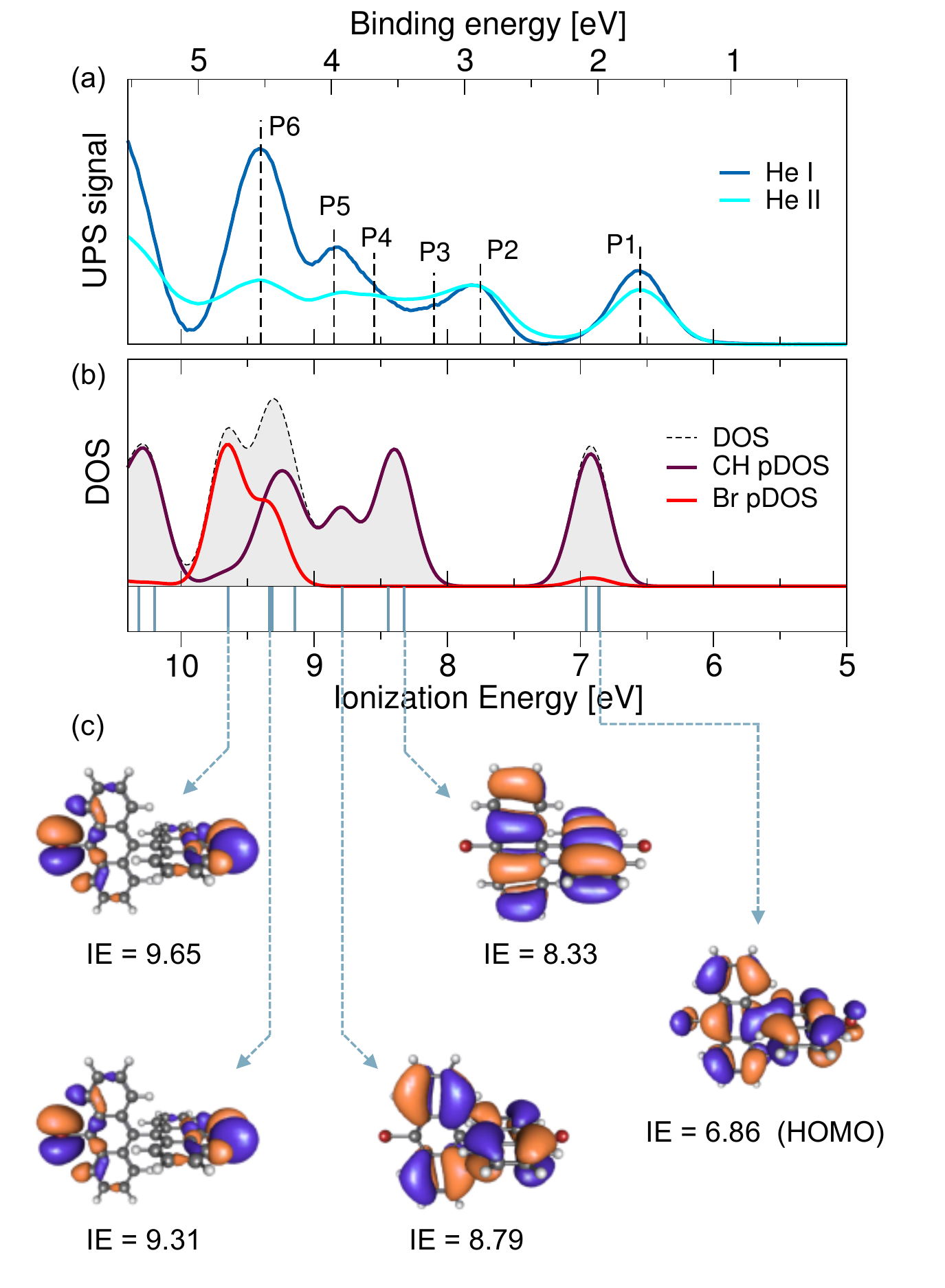} 
\fi
\caption{(a) Normal emission UPS spectra of DBBA-TF deposited on the Au(110) surface at RT, measured with He-I (21.218 eV, blue) and He-II (40.814 eV, cyano) photon energies. The work function of the system ($\Phi$=4.87 eV) has to be added to the binding energies in order to define the ionization energies (IE).
Spectra are normalised to the intensity of the peak P2. 
(b) Total and projected (on CH and Br atoms) DOS computed by DFT (CAM-B3LYP functional) for an isolated (gas phase) DBBA molecule, plot with respect to the vacuum level (IE).
The Gaussian broadening of the molecular levels has been set to 0.3 eV. 
(c) Calculated isosurfaces of constant density for selected DBBA orbitals; orbital ionization energies are also reported.  
\label{Fig_DBBA_UPS}}
\end{figure}

The photoemission intensity excited by He-I photons and acquired at increasing DBBA coverages, is shown in Fig.~\ref{EDC_coverage_dependence_RT}c. As the coverage increases, the Au surface states are quenched and the molecular states become more and more intense, until they dominate the spectrum for the DBBA-TF.  
For the latter, six non-dispersive molecular peaks (P1-P6) can be identified between the Fermi level (E$_\text{F}$) and 5~eV binding energy (BE), as shown in Fig.~\ref{Fig_DBBA_UPS}a. 
The spectral features in this energy range are expected to be dominated by the $\pi$-states of the conjugated molecule, while $\sigma$-states lie at larger binding energies, in agreement with the existing literature on analogous aromatic molecules  \cite{Corradini_surf_science2003,Yamauchi_1998,Hunger_JPCB_2006,Fujisawa_JACS_1986,Potts1980,Baldacchini2007,Alagia_JCP2005}.

The spatial localization of the electronic states on the C and Br atoms can be experimentally unraveled by exciting the photoelectrons with different photon energies. In fact, according to the photo-ionization cross section values for Br $4p$ \cite{Yeh_1985,Yeh_1993}, the signal from an orbital with a charge density located on Br should be enhanced by He-I photons (21.218 eV) as compared to He-II (40.814 eV). 
The spectra acquired with He-I and He-II radiation were normalized to the intensity of the P2 peak (see Fig.~\ref{Fig_DBBA_UPS}a), which is predicted by DFT to be fully localized on the C aromatic rings (see Fig.~\ref{Fig_DBBA_UPS}b-c, and description below). This procedure removes cross-section differences of the C $2p$ atomic levels while highlighting those due to Br atoms, thus allowing the identification of Br contribution to the different molecular peaks. 
We find that the peaks from P1 to P4 have a dominant $\pi$ character, even though a small Br $4p$ contribution can be associated with P1 (superposition of HOMO and HOMO-1, according to DFT calculations, see Fig.~\ref{Fig_DBBA_UPS}b-c). On the contrary, an intensity increase of the He-I signal with respect to the He-II one is observed for peaks P5 and P6, indicating a prominent Br contribution. 
The significant Br contribution observed in these  molecular peaks demonstrates the integrity of the majority of DBBA molecules upon adsorption on Au(110) at RT, in agreement with previous photoemission core-level data \cite{massimi_JPCC_2015}.

\begin{table}
\centering
\begin{tabular}{l lll l}
\toprule
\\[-5pt]
      [eV]      & IP$_{\Delta\text{-SCF}}$ & & HOMO   & LUMO  \\[7pt]
\midrule
\\[-5pt]
LDA             & 6.90 (6.90)  & & 5.32 (5.30)    & 3.21 (3.18)    \\
PBE             & 6.67 (6.69)  & & 5.10 (5.14)    & 2.97 (3.00)    \\
B3LYP           & 6.75         & & 5.61           & 2.34     \\
CAM-B3LYP       & 7.14         & & 6.86           & 1.35     \\
PBE0            & 6.92 (6.96)  & & 5.88 (5.90)    & 2.32 (2.31)    \\
BHH             & 6.95         & & 6.44           & 1.43     \\[7pt]
\hline
exp             &              & & 6.57   &      \\[7pt]
\bottomrule
\end{tabular}
\caption{Ionization potentials computed for DBBA using a localized basis set (aug-cc-pVTZ basis) following the $\Delta$-SCF procedure (see text), for different functionals. Absolute values (negative eigenvalues) of HOMO and LUMO are also reported. Equivalent calculations done by using Quantum ESPRESSO are shown in parentheses for comparison.
The ionization energy of the HOMO measured by UPS for a DBBA-TF is reported (exp) for reference.}
\label{tab:theo_IP}
\end{table}

\begin{figure}
\centering
\iffigures
    \includegraphics[width=0.5\textwidth]{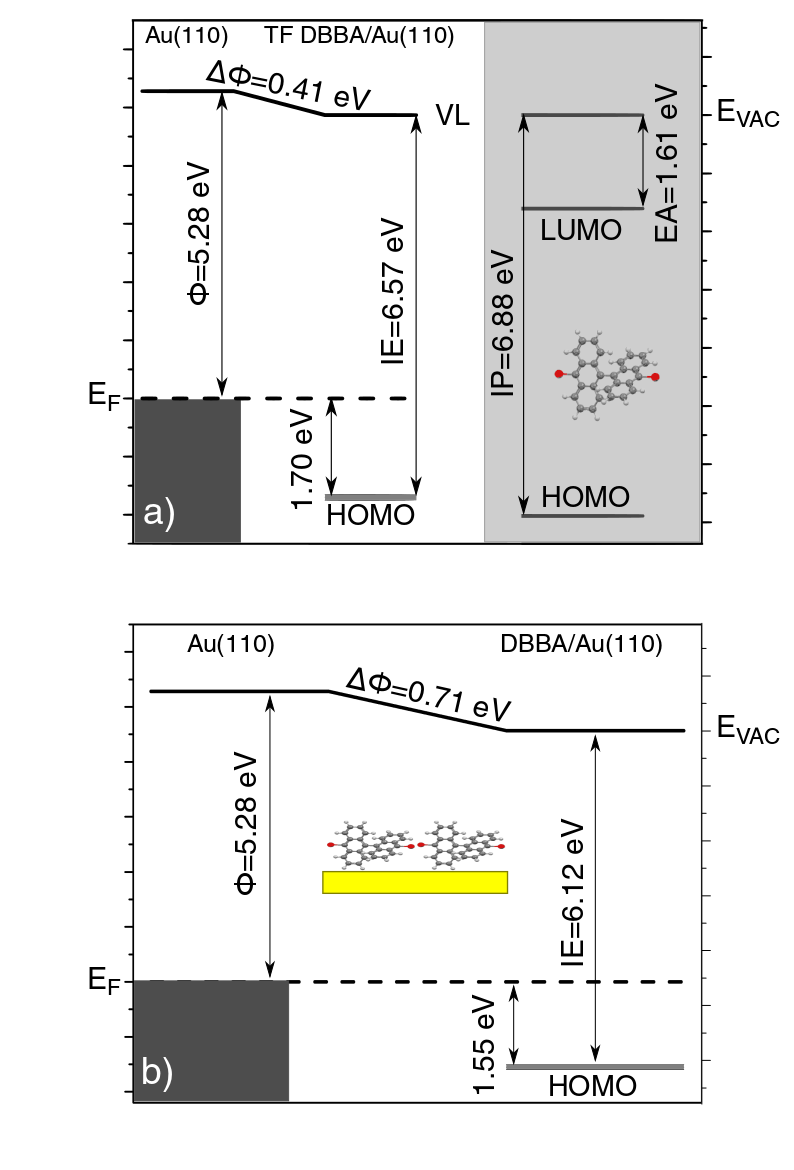}
\fi
\caption{Energy level diagrams. (a) Clean Au(110) surface (left), DBBA-TF on Au(110) (middle), and theoretical results (IP and EA from the Slater's-1/2 method at the LDA level) for gas-phase DBBA molecules (right). (b) DBBA-SL on Au(110) with respect to the clean Au surface.
\label{energy_level_diagram}}
\end{figure}

In order to support the analysis based on UPS data, we have computed by DFT the density of states (DOS) and the molecular orbitals for the isolated DBBA molecule, which are shown in Fig.~\ref{Fig_DBBA_UPS}b-c.
The DBBA ionization potential (IP) was computed by using the $\Delta$-SCF procedure (i.e. as the difference of the ground state energy of the neutral and the cationic species), which is expected to provide very good estimates of the vertical ionization energies (IE)~\cite{dabo+10prb,borg+14prb}. The results obtained by using several exchange and correlation functionals are summarized in Table \ref{tab:theo_IP}. Already at the LDA level, the IP from $\Delta$-SCF (6.9~eV) compares fairly well with the UPS data for the DBBA-TF, which gives an IP value of 6.57~eV (see Tab.~\ref{tab:theo_IP} and Fig. \ref{Fig_DBBA_UPS}). Note that the difference of $\sim$0.3 eV found here between theory and experiment is not likely to be totally ascribed to the theory accuracy. In fact,  the IP value (calculated or measured) for gas-phase molecules should be compared with caution to  that  found  for  a  corresponding  TF,  the  latter  being  usually  smaller  due  to  solid  state effects (screening, surface polarization, intermolecular interaction). 
Taking pentacene as an example, previous experimental and theoretical results show a residual energy shift between TF and gas-phase IP's as large as 0.55 eV~\cite{Baldacchini2005phd,Baldacchini2006,Baldacchini2007,FerrettiPRL2007}. Instead, the agreement of the pentacene gas-phase IP from UPS (6.4-6.6 eV from Refs. ~\cite{Baldacchini2005phd,Baldacchini2006,Baldacchini2007,Schmidt1977,Coropceanu2002}) and the IP computed using the  $\Delta$-SCF method (6.40 eV, computed here for isolated pentacene at the LDA level) is remarkable. 
Concerning DBBA, very similar values to those obtained by $\Delta$-SCF are found by using the Slater's-1/2 approach\cite{Slater1974} (i.e. HOMO and LUMO energies are computed as eigenvalues at half-occupation), which gives a value of 6.88 eV for the IP, and allows us to compute also the electron affinity (EA = 1.61 eV), as shown in the energy level diagram of Fig.~\ref{energy_level_diagram}a.
On the contrary, much larger errors than for $\Delta$-SCF or Slater's-1/2 are found when estimating the IP directly from the HOMO eigenvalues, as shown in Tab.~\ref{tab:theo_IP}. 
The discrepancy between total energy differences and eigenvalues, i.e. the failure of the Koopmans' condition~\cite{dabo+10prb,borg+14prb}, has been recently used to construct functional corrections aiming at piece-wise linearity~\cite{borg+14prb,nguy+15prl}. In this regard, hybrid functionals perform better than LDA and GGA, the best case being that of CAM-B3LYP, with only a 0.3 eV difference between $\Delta$-SCF and HOMO energies. For this reason, we have chosen the DOS computed with CAM-B3LYP as a reference to compare with experimental data.
Indeed, for the whole energy range shown in Fig.~\ref{Fig_DBBA_UPS}, there is a very good agreement between the CAM-B3LYP DOS and the UPS spectra, in terms of both ionization energies and Br-contribution of the molecular features. In particular, the calculated projected DOS shows the same Br contributions as the experimental P5 and P6 peaks in the IE energy range 9-10~eV.
Such a good correspondence further confirms that the molecular states of the DBBA-TF can be assigned to the DOS of intact and weakly interacting DBBA molecules. 

The molecular level distribution resulting from the above analysis constitutes a useful reference to understand the evolution of the electronic states occurring in a DBBA-SL adsorbed on Au(110). To summarize our findings and make the comparison easier, we report in Fig.~\ref{energy_level_diagram} the energy level diagrams resulting from UPS data for both the DBBA-TF (as compared to DFT-LDA calculations for gas-phase DBBA) and the DBBA-SL deposited on Au(110) at RT, which is further discussed below.

\subsection{DBBA and polyanthryl on Au(110)}

\begin{figure}
\centering
\iffigures
   \includegraphics[width=0.5\textwidth]{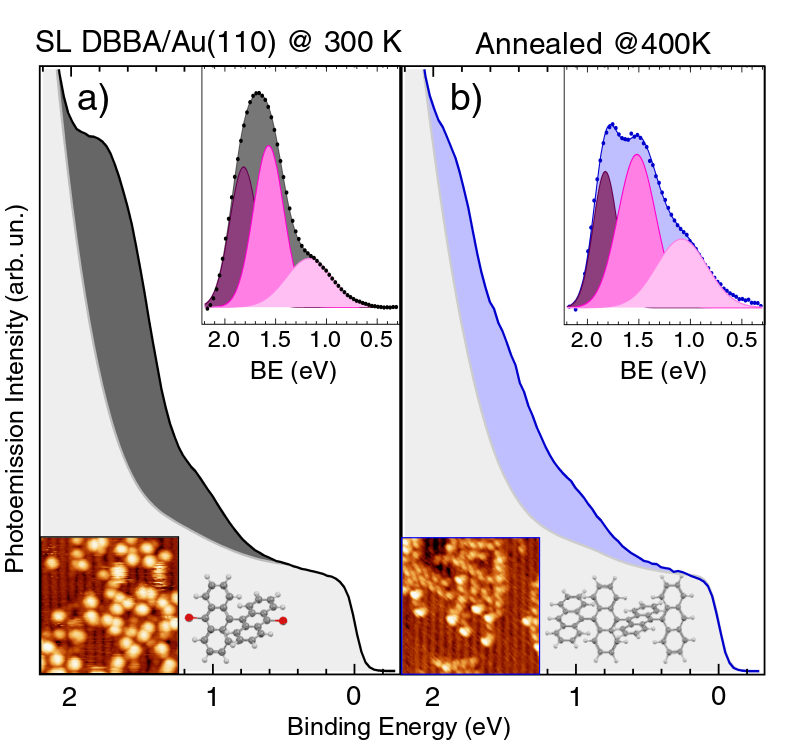}
\fi
\caption{Normal emission UPS spectra in the low BE region for (a) DBBA-SL deposited on Au(110) kept at RT and (b) after annealing to 400 K. A background (grey area, Fermi edge and spline) has been subtracted to obtain the top insets. Data taken along the direction at 20$^{\circ}$ with respect to the Au[001] azimuthal direction. In the bottom insets, scanning tunneling microscopy images\cite{massimi_JPCC_2015} (dimensions: 13.5$\times$13.5 nm$^2$) and sketches of the molecular systems in the two conditions. 
\label{SL_poly_vs_DBBA_vs_gold}}
\end{figure}

The adsorption of a DBBA-SL on a Au(110) surface at RT is the first step of a thermally activated process leading to dehalogenation, molecular polymerization and, eventually, to the synthesis of GNRs $via$ cyclo-dehydrogenation\cite{massimi_JPCC_2015}. The evolution of the molecular states of a DBBA-SL is analyzed by means of photoemission spectroscopy in the low BE region close to the Fermi level (see  Fig.~\ref{SL_poly_vs_DBBA_vs_gold}). In the two panels we plot the UPS signal corresponding to (a) DBBA-SL as deposited at RT and (b) after annealing at 400 K for 10 min. Background-subtracted spectra are displayed in the top insets. 

For the DBBA-SL deposited at RT (top inset of Fig.~\ref{SL_poly_vs_DBBA_vs_gold}a), we can identify the Au(110) surface state at 1.83 eV BE \cite{Hansen_SS_1989} (purple curve in the inset), being strongly reduced in intensity with respect to clean Au (S peak in Fig. \ref{EDC_coverage_dependence_RT}c), as well as a slightly increased spectral density in the lower BE region (0-2 eV). In particular, we observe two main features in this range, one at 1.55 eV and the second at 1.15 eV.
The peak at 1.55 eV BE (magenta curve) can be assigned to the HOMO state of the DBBA molecules on Au(110), which is found at slightly lower BE than the HOMO measured for the condensed DBBA-TF (1.70 eV BE). The energy lowering of molecular states in the vicinity of a metal surface can be explained by the photo-hole screening due to the metal\cite{Neaton2006}, as also observed for other aromatic molecules adsorbed on Au(110)  \cite{Gargiani_PRB_2010,Massimi_JCP_2014,Baldacchini2006}. 
The second feature at 1.15 eV BE (pink curve) can instead be associated with the presence of substrate-induced dehalogenated DBBA (i.e. bianthracene, BA), with Br-cleaved C atoms interacting with the Au surface to saturate their bonds. As previously shown by XPS spectroscopy~\cite{massimi_JPCC_2015}, most of the DBBA molecules absorb intact on the Au(110) surface at RT. 
However, a sensible fraction of them is already dehalogenated at RT, as confirmed by the appearance of a broad C-1s component at low BE in the XPS spectra~\cite{massimi_JPCC_2015}. The presence of C-metal bonds have been previously reported for unsaturated adsorbates in similar systems \cite{Simonov_ACSnano2015, Simonov_correction_JPCC_2015, Lacovig_PRL_2009_Graphene_nanoislands}. No additional photoemission features are visible in the energy region up to the Fermi level, allowing us to rule out the presence of a net charge transfer or interface states associated to partially-filled LUMO-derived states, as reported for other $\pi$-conjugated molecular systems on metal surfaces \cite{Heimel_Nat_Chem_2013,Annese_PRB2008,FerrettiPRL2007,Baldacchini2007}. 
The absence of interacting states can be explained by the non-planar geometry of the DBBA molecules adsorbed on Au(110), due to the steric hindrance between the inner hydrogen atoms of the molecules. Once a net charge transfer between the molecule and the substrate is excluded, one can reconsider the work function variations observed at increasing DBBA coverages. The energy level diagram reported in Fig.~\ref{energy_level_diagram}b shows a lowering of 0.71 eV of the work function at the completion of the SL (see also Fig.~\ref{EDC_coverage_dependence_RT}b). 
The observed decrease of the work function indicates the creation of interfacial vertical dipoles~\cite{Ishii1999,Braun2009,DellaPia_ACSnano2014,TombaACS2010}, possibly associated with the so-called pillow  (or cushion) effect~\cite{WitteAPL2005}. This dipole is created whenever a molecule or an atom are deposited on a metal surface and is due to the Pauli repulsion between the tail of the electronic cloud of the metal and the electrons of the adsorbates. The existing repulsion effectively pushes back the protruding tail thus leading to the formation of a dipole localized at the position of the molecule. Vertical dipoles could also arise due to the distorted adsorption conformation of DBBA \cite{massimi_JPCC_2015} and/or as a result of depolarization effects typical of electron-rich aromatic molecules such as DBBA~\cite{Topping_1927,LacherJACS2011,Fraxedas2011}. 
After deposition at RT, the DBBA-SL has been further annealed at 400 K. In this condition, the dehalogenation reaction proceeds and the adsorbed monomers start to diffuse and to coalesce into polyanthryl (PA) chains, with an average preferential orientation of $\pm$20$^{\circ}$ with respect to the Au[001] azimuthal direction, as elucidated by previous structural and spectroscopic investigations \cite{massimi_JPCC_2015}. 
However, at variance with the DBBA adsorbed on Au(111) or Cu(111) surfaces \cite{Simonov_JPCC_2014,Cai_nature_2010}, the polymer length is shorter and their density is lower, due to the reduced molecular mobility on the corrugated Au(110) surface \cite{massimi_JPCC_2015}. 

As compared to DBBA-SL at RT, the UPS spectrum after annealing (Fig. \ref{SL_poly_vs_DBBA_vs_gold}b) shows only minor changes, i.e. a slight redistribution of the intensity of the main peaks with respect to DBBA and a small shift (100 meV) of the peak associated with the C-Au interaction (1.15 eV BE, pink curve in panel a). While a similar low-BE peak, associated with C-Au bonds, could be expected from unsaturated PA edges (with small changes possibly associated with the conformational difference with respect to single BA molecules), the absence of major changes with respect to the position of the DBBA HOMO would imply an almost size-independent electronic structure for the PA chains.

\begin{figure}
\iffigures
   \centering
   \begin{tabular}{m{0.5\textwidth} m{0.13\textwidth}}
   \includegraphics[width=0.5\textwidth]{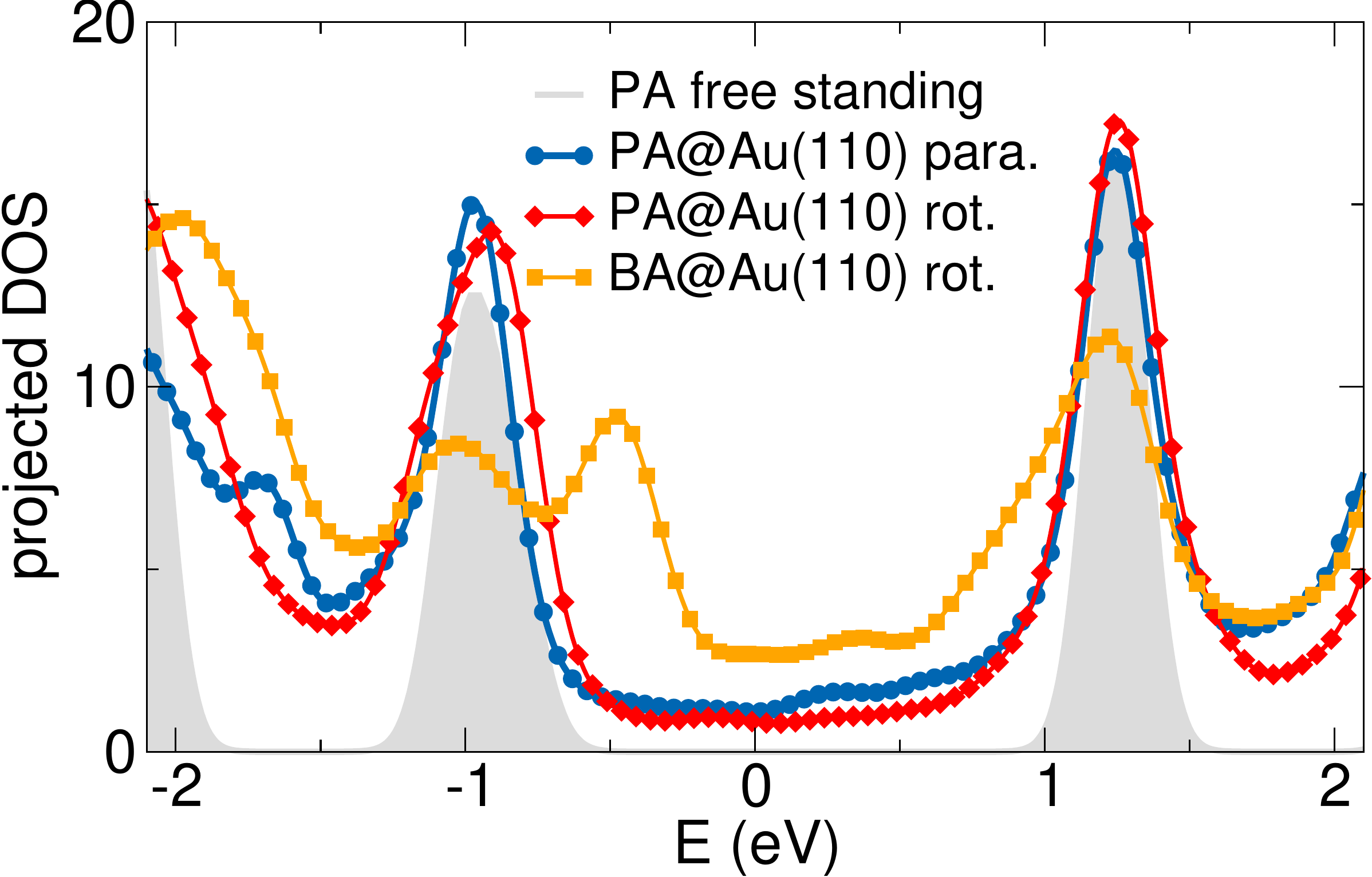}&
   \includegraphics[width=0.1\textwidth]{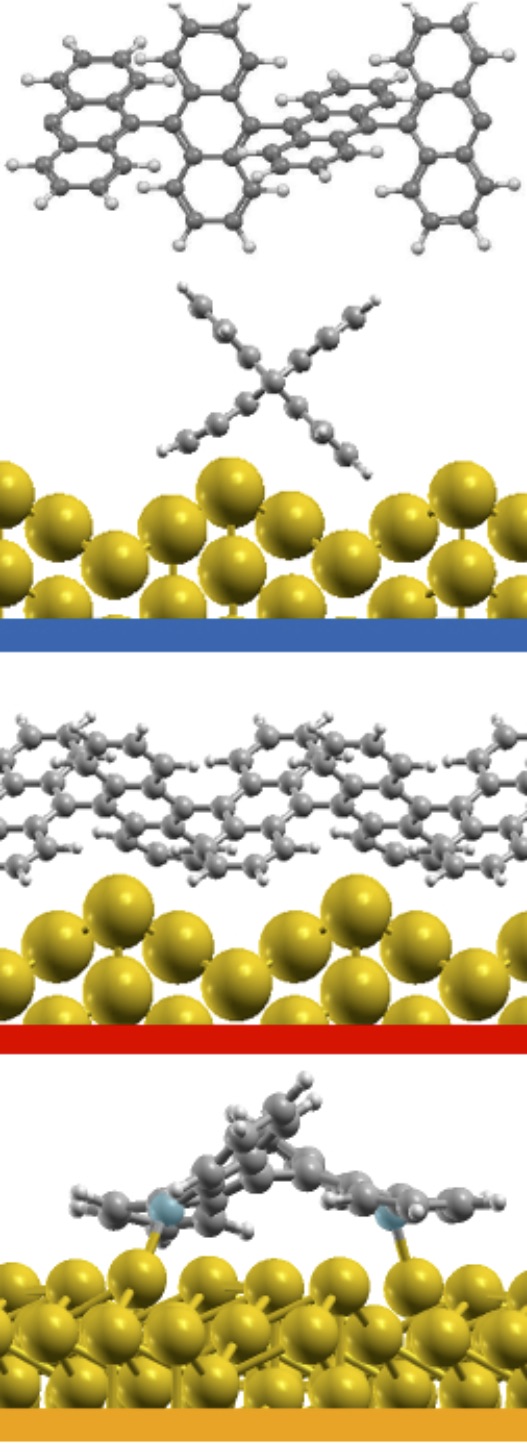}\vspace{12pt}
   \end{tabular}\\
   \includegraphics[height=0.45\textwidth]{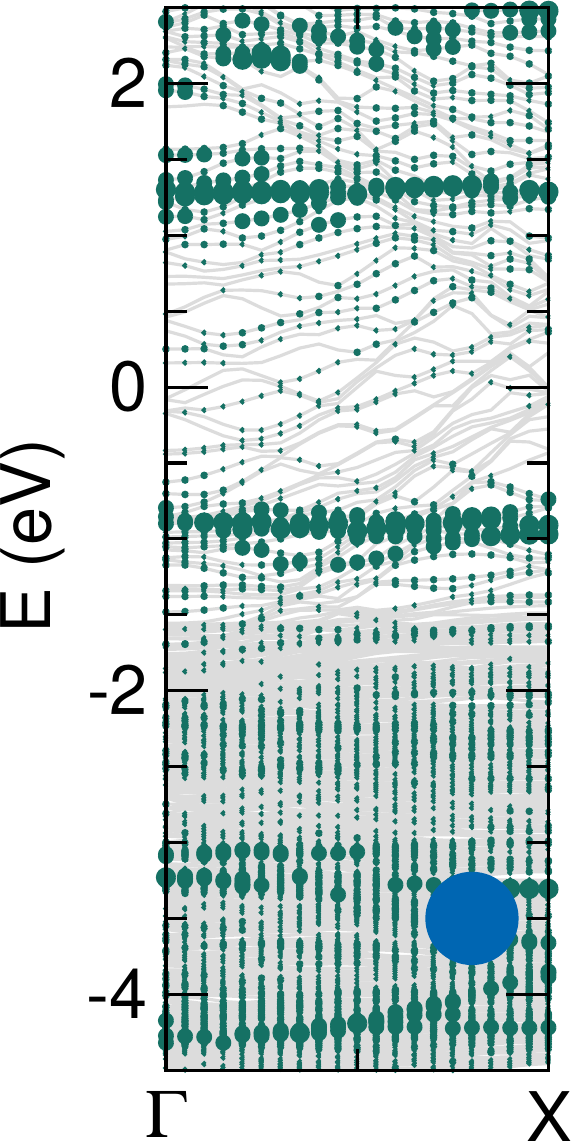}\hspace{0.03\textwidth}
   \includegraphics[height=0.45\textwidth]{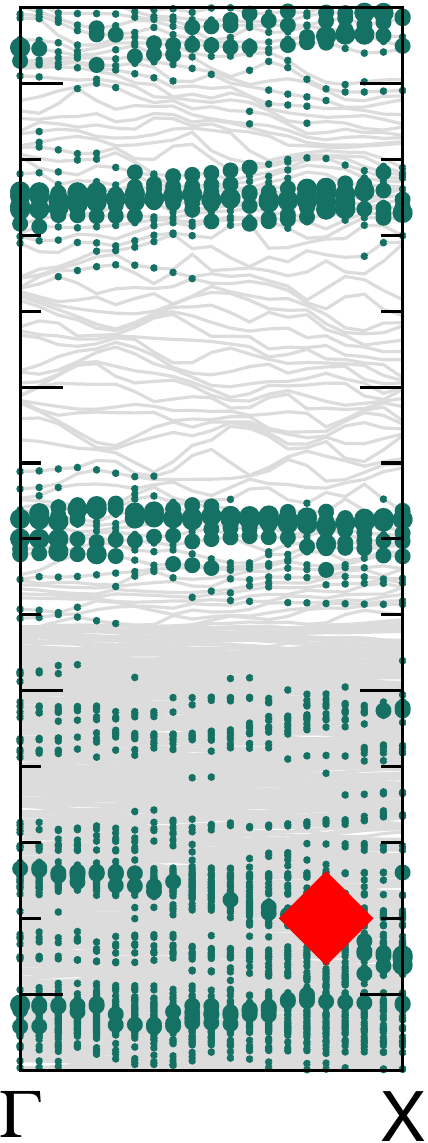}\hspace{0.03\textwidth}
   \includegraphics[height=0.45\textwidth]{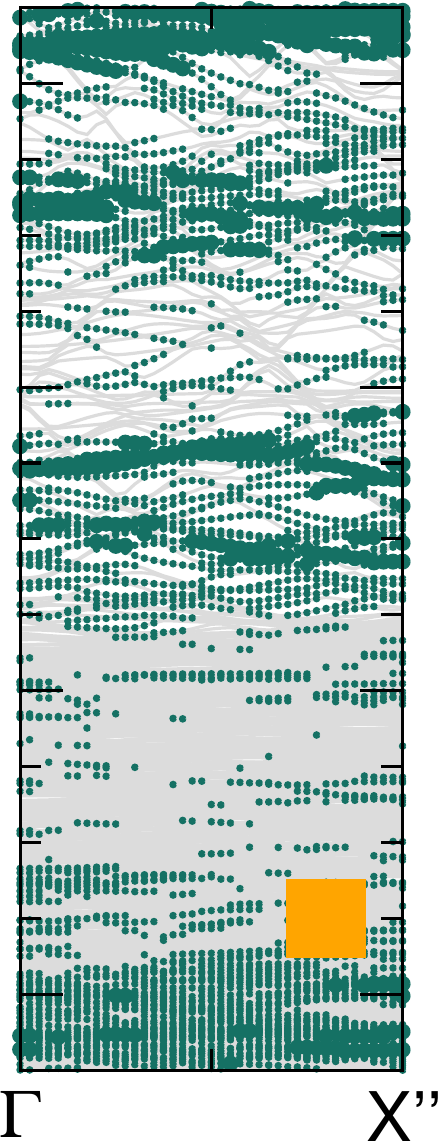}\hspace{0.03\textwidth}
\fi
\caption{DOS (upper left panel) and band structure (lower panels), projected on polyanthryl (PA) when adsorbed on a 1$\times$2 Au(110) reconstructed surface, both parallel to the Au rows (blue circles) and rotated by 70$^\circ$ with respect to the same rows (red diamonds) and projected on dehalogenated bi-anthracene (yellow squares). 
The DOS of free standing PA is also reported as a reference (grey shaded area) and its band structure is shown in Fig. S2 of the SI.
The studied configurations are shown in the upper right panel. In the band structure plots the size of the green dots is  larger for states with larger projection on PA or BA. The grey lines represent the Au states. All calculations were performed within DFT/LDA. 
\label{poly_mol_DFTbands}}
\end{figure}

In order to validate these hypotheses, we firstly investigate the length dependence of the electronic structure of PA oligomers, by comparing the gas-phase DOS of several oligomers constituted by 1 to 5 units of DBBA. 
For all the considered levels of theory (see Methods), the oligomers of different lengths show a very similar DOS, and no change in the HOMO position (see Supplementary Information). This implies that the IP does not depend on the oligomer length and is expected to be the same for both DBBA and PA, in agreement with the experimental findings (see Fig. \ref{SL_poly_vs_DBBA_vs_gold}). 
The reason for this similarity between single molecules and oligomers probably lies in the broken conjugation between the anthracene units, which also results in the non-dispersive bands of the PA polymer discussed below.

We next consider the effect of the Au substrate, by comparing the DFT-LDA (projected) DOS for bianthracene (BA) and PA adsorbed on a Au(110) 1$\times$2 reconstructed surface, as shown in the upper panel of Fig.~\ref{poly_mol_DFTbands}.
For the polymer, we have considered two different configurations, in which PA is either aligned along the Au(110) channels or bridging the channels according to the orientation observed experimentally (70$^\circ$ with respect to the channels, see discussion above and Ref.~\cite{massimi_JPCC_2015}). The two configurations do not lead to any significant change in the DOS, indicating a similar, weak interaction with the Au surface in both cases. This is also confirmed by the similarity with the DOS of free-standing PA (grey area, aligned with the molecular features at -0.9 eV), as well as by the absence of spectral features in the gap energy region. 
This is even more clear in the lower panels of Fig.~\ref{poly_mol_DFTbands}, where the bands projected on PA are shown. Here larger dots correspond to larger projections of the system wavefunctions on the C atoms. The projected bands of PA are found to be mostly non-dispersive, and very similar for the rotated and the aligned configurations. The band structures do not show a significant hybridization with the Au states, shown here as grey lines. 

Comparing the DOS for PA and BA, we find that the peak corresponding to the PA HOMO, located at -0.9~eV, splits into two contributions of about the same intensity in the case of BA, now located at -1.0 and -0.5~eV. The DOS projected on the different C atoms (see Supporting Information) shows that the main contribution to the new peak at -0.5~eV is from the two carbon atoms bound to the Au surface rows (halogenated in the intact molecule). A similar effect is expected for oligo-anthryl units, where unsaturated bonds are also present. This is however not included in our calculations, since PA is simulated as an extended, periodic system. 

To conclude, the theoretical DOS predicts: ($i$) the appearance of a state that can be associated with the interaction of BA molecules with the underlying metal atoms, favored by the adsorption geometry ($\pm$20$^{\circ}$ with respect to Au[001] direction); ($ii$) an almost size-independent HOMO for the PA oligomers, which are also ($iii$) weakly interacting with the underlying Au substrate, as demonstrated for extended PA. All of this appears to be in very good agreement with experimental observations from UPS.

\subsection{GNRs on Au(110)}

\begin{figure}
\iffigures
   \centering
   \begin{tabular}{m{0.5\textwidth} m{0.13\textwidth}}
   \includegraphics[width=0.5\textwidth]{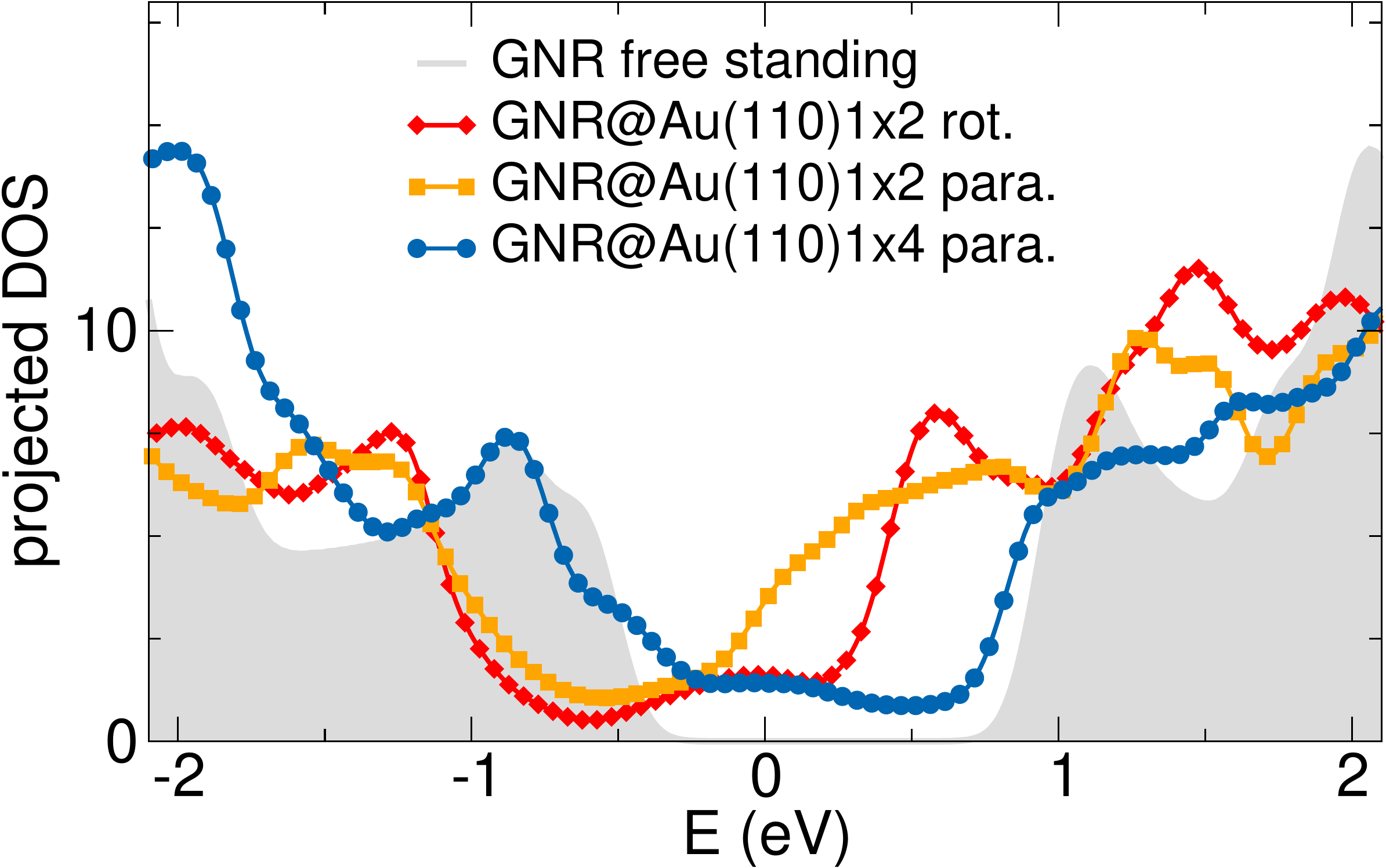}&
   \includegraphics[width=0.1\textwidth]{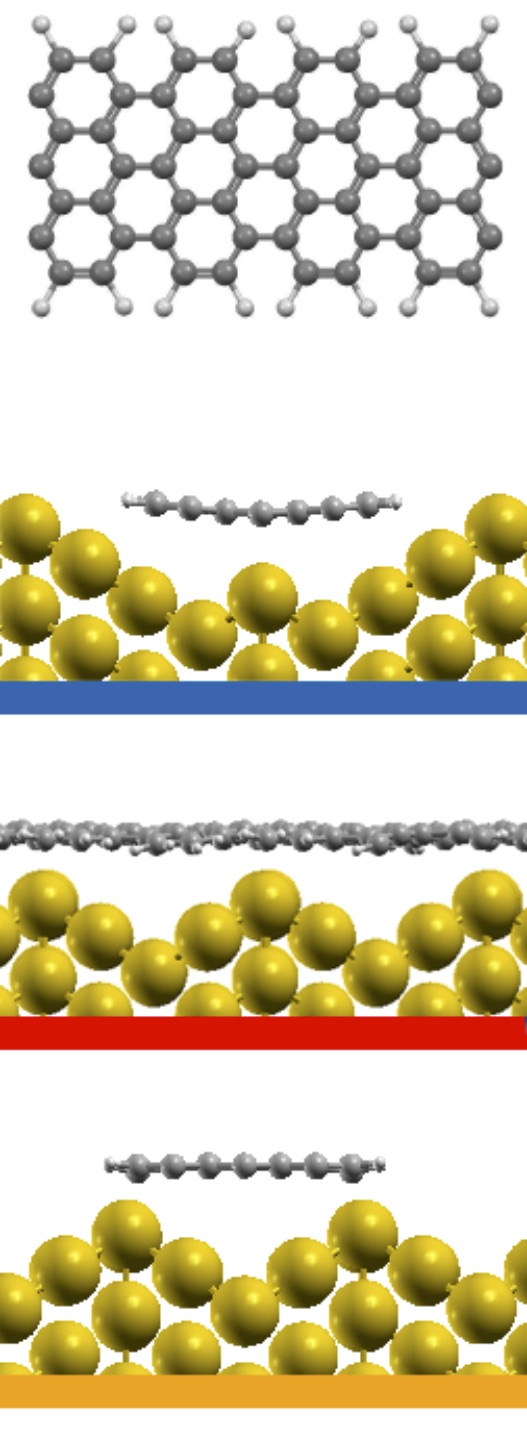}\vspace{12pt}
   \end{tabular}\\
   \includegraphics[height=0.45\textwidth]{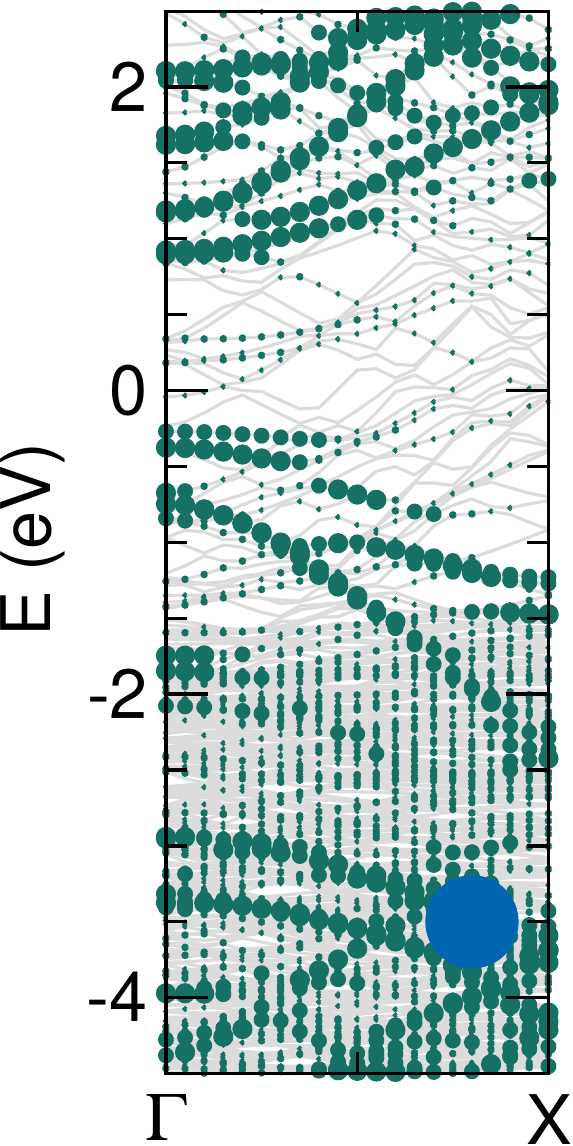}\hspace{0.03\textwidth}
   \includegraphics[height=0.45\textwidth]{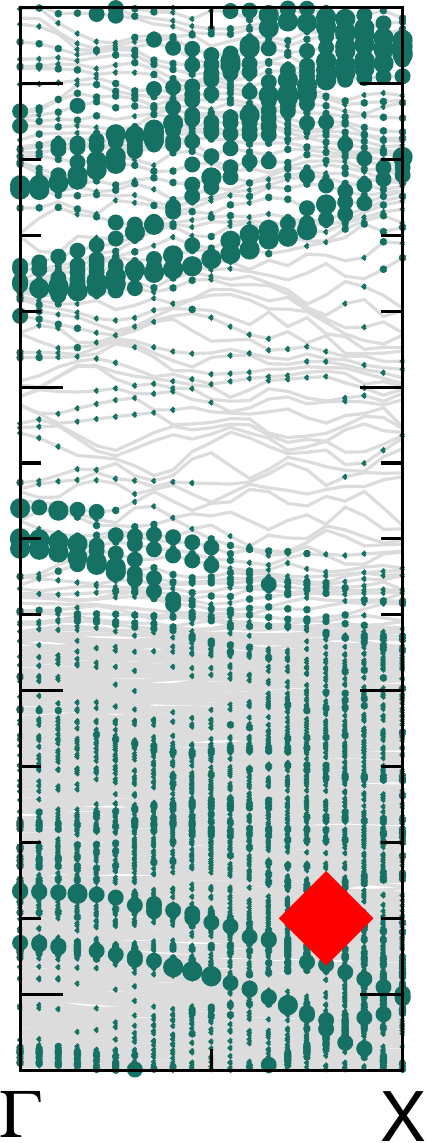}\hspace{0.03\textwidth}
   \includegraphics[height=0.45\textwidth]{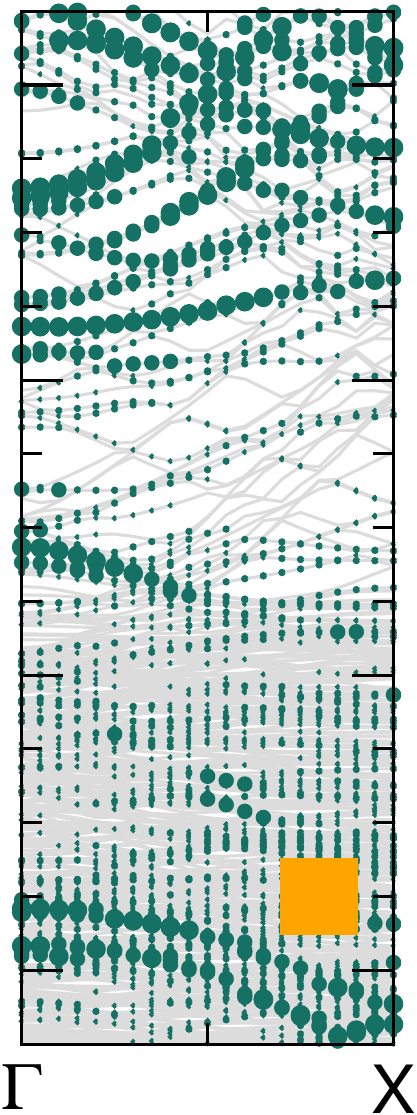}\hspace{0.03\textwidth}
\fi
\caption{DOS (upper panels) and band structure (lower panels) projected on graphene nanoribbon (GNR) in different conditions: free standing (grey shaded area), adsorbed on a 1$\times$2 Au(110) (red diamonds and orange squares for GNR aligned or rotated by 70$^\circ$ with respect to the  [$1\bar{1}0$] direction respectively), and on a 1$\times$4 Au(110)  (blue circles). The studied configurations are shown in the upper right panel. In the band structure plots the size of the green dots is  larger for states with larger projection on GNR. All calculations are done within DFT/LDA. The band structure of free standing GNR is shown in Fig. S2 of the SI.
\label{GNR_DFTbands}}
\end{figure}

Different chemical paths have been identified~\cite{massimi_JPCC_2015} to achieve sub-nanometer wide GNRs on Au(110). In the first synthetic route, after PA chains have been formed at 400 K, further annealing at 700 K spurs the dehydrogenation process and the related flattening of the C network towards the formation of short GNR strands, oriented as the original PA chains.    
The spectral density (not shown here) of these GNRs, also characterized in Ref.~\cite{Batra_Chem_Science2014}, is redistributed in the low binding energy region close to the Fermi level. However, neither well-defined peaks nor dispersive electronic states are observed, as expected upon formation of extended structures. On the contrary, GNR strands oriented along the 4-fold reconstructed gold rows have been obtained following the second synthetic procedure described in Ref.~\cite{massimi_JPCC_2015}. 
Briefly, the gold substrate is kept at 470 K during deposition, and both dehalogenation and dehydrogenation take place before any polymerization. By increasing the molecular coverage up to 0.5 SL, bisanthene molecules (originating by flattening bianthracene units) form ordered chains along the [$1\bar{1}0$] direction, either bridging two adjacent rows of the pristine 1$\times$2 Au reconstruction or inside rearranged 1$\times$3 gold channels. At completion of the first SL, the Au surface further rearranges into a 1$\times$4 symmetry, with elongated nanostructures aligned along the channels ([$1\bar{1}0$] direction). The scanning tunneling microscopy images suggest the coexistence of dehalogenated and dehydrogenated single molecules with GNRs segments~\cite{massimi_JPCC_2015}. 
The bisanthene structures and the short GNR strands formed on the  1$\times$4  reconstructed channels lead to a broadened spectral density in the energy region close to the Fermi level (see Supplementary Information), with similar features to those observed for GNRs obtained by following the first synthetic path. In both cases, the presence of a limited amount of short GNR strands with different adsorption configurations and hence different band structure (see theoretical findings below) hinders the experimental detection of dispersive electronic states. 
In analogy with the case of PA, DFT calculations have been performed for adsorbed GNRs to clarify the experimental results and to study the role of their interaction with the Au(110) substrate.
DFT calculations for GNRs adsorbed on Au(110) have been performed considering three different configurations: ($i$) GNR adsorbed on Au(110) 1$\times$2 aligned along the Au rows; ($ii$)  GNR adsorbed on Au(110) 1$\times$2 along the direction which forms an angle of 20$^\circ$ with the Au[001] direction; ($iii$) GNR adsorbed on a 1$\times$4 reconstructed surface inside the Au(110) rows.
This allows us to investigate the effect of the GNR orientation on the interface electronic structure.
In Fig. \ref{GNR_DFTbands} we present the DOS projected on the GNR for the three cases.
The DOS of a free-standing GNR is shown as a reference, by aligning it to configuration ($iii$), which is the one observed experimentally.

When comparing to PA (Fig.~\ref{poly_mol_DFTbands}), the DOS of GNRs appears broadened and less sharp, in agreement with photoemission results reported in the supporting information, indicating a dispersion of the bands (clearly visible in the bottom panels of Fig.~\ref{GNR_DFTbands}). This is already the case when considering the corresponding free-standing systems and reflects the fact that full $\pi$-conjugation is established in the flat GNR, contrary to what happens for PA.
Moreover, also at variance with PA, the DOS for different GNR configurations on Au(110) are significantly different, and distinct from the DOS computed for the free-standing GNR. This indicates a stronger effect of the underlying substrate, that can be ultimately attributed to the flat structure of the molecules. 

The DOS computed for the rotated GNR bridging the Au rows on Au(110) 1$\times$2 is the closest to the free-standing case, which agrees with the fact that the average distance to the surface is larger, and therefore the GNR-Au interaction weaker than for the other two configurations. Instead, the GNRs adsorbed along the 1$\times$2 Au channels show a non-negligible hybridization with the Au states around the Fermi energy that is not present in the case of the rotated GNR. 
Similarly, the GNRs aligned along the 1$\times$4 channels show some hybridization-related features at about 0.5 eV below the Fermi level (together with an overall shift with respect to the Fermi energy, as compared to the other two cases). 
This is a consequence of the fact that GNRs aligned along the channels bridge two rows of prominent Au atoms all along their length.


\section{Conclusions}
%
The electronic structure of different phases of DBBA deposited on Au(110) has been followed by means of photoemission spectroscopy and ab-initio DFT calculations, allowing us to determine the alignment of molecular energy levels at the hybrid interface during the reactions steps leading to DBBA polymerization and, eventually, GNR formation.

Comparing UPS data for DBBA deposited on Au(110) at room temperature (thin film) and DFT calculations, it is possible to identify the contribution of C and Br to the different molecular peaks, confirming the presence of pristine molecules on the surface. The ionization energy computed for the DBBA HOMO in gas phase (6.9~eV) is in very good agreement with the value measured for the thin film (6.57~eV).
Concerning the DBBA/Au interface, the dehalogenation of the molecules gives rise to the formation of C-Au bonds, which leads to a redistribution of the spectral density close to the Fermi level. This is seen already in the RT UPS spectra as a peak at 1.15~eV below the Fermi energy, which persists after the annealing at 400~K. Consistently, similar features are found in the computed density of states, as a splitting of the HOMO DBBA states when the C-Au bond is formed. 
This scenario supports the idea that unsaturated carbon bonds at the molecular edges may play an important role in the reactivity with the metal and hence in the reaction steps leading to graphene nanostructures. 

The electronic structure computed for polyanthryl adsorbed on Au does not differ significantly from that of the free-standing polymer. On the contrary, the GNR band structure is more sensitive to the presence of the substrate, showing clear differences in the valence spectral density close to the Fermi level when changing orientation and surface reconstruction, a sign of a stronger interaction with Au than in the case of polyanthryl. 
Ultimately, such a stronger interaction can be ascribed to the flat structure of the GNRs that, besides leading to larger dispersion in the GNR DOS, also triggers an adsorption geometry characterized by shorter distances with the metal substrate. 

\section{Methods}
\subsection{Experimental methods}
Experiments were performed in a ultra-high-vacuum (UHV) chamber with base pressure in the low 10$^{-10}$ mbar range, at the Lotus Surface Science laboratory in Rome. Ultraviolet photoemission spectroscopy (UPS) experiments were performed by using He-I (21.218 eV) and He-II (40.814 eV) radiation provided by a Scienta VUV-5050 monochromatic source. Photoelectrons were collected through a Scienta SES-200 analyzer with an angular acceptance of $\pm 8^{\circ}$ and a resolution of 16 meV. Normal emission data have been integrated over an overall 12$^{\circ}$-wide angular window. 
The secondary electrons cut-off (SECO) was measured by applying a voltage ranging from -2.40 to -3.00 V to the sample with respect to ground. Work function values were calculated by subtracting the difference between the SECO and the Fermi energy from the photon energy. The intersection between a linear extrapolation of the actual cut-off and the horizontal, zero photoemission intensity axis was used as SECO value. The uncertainty in the work function value, equal to 0.02 eV, is linked to the energy resolution of the analyser and to error propagation. 

The Au(110) surface was cleaned by cycles of sputtering (800 eV Ar$^+$ at a pressure of 10$^{-6}$ mbar) and annealing (720 K), followed by a final sputtering cycle with 400 eV Ar$^+$ ions at a pressure of 10$^{-6}$ mbar, performed on the sample held at 570 K.
Sample long-range order and cleanness were checked by means of low energy electron diffraction (LEED) 
(finding a clear 1$\times$2 reconstruction) and by surface states analysis by means of UPS.

DBBA molecules (commercially purchased from Richest Group Ltd) were degassed in UHV for several hours at 400 K, and then sublimated by means of a home-made resistively heated quartz crucible at 420 K and 10$^{-9}$ mbar. 
A quartz crystal microbalance (QCM) was used to calibrate the molecular coverage, determined by multiplying the deposition rate measured by the QCM and the deposition time. The coverage relative error on this nominal calibration (correct under the assumption of a constant sticking coefficient and no material desorption) has been estimated to be 15 \% by comparing different sets of experiments. 
In the case of deposition on a Au(110) kept at 300 K, we define the nominal single-layer coverage as the point where the work function reaches its minimum value when plotted against coverage (see Fig.~\ref{EDC_coverage_dependence_RT}b). We point out that this is an upper limit because the molecules may have a conformational dipole and hence the work function is expected to slightly decrease after the single layer (SL) completion. 
An estimation of the coverage of the thin film can be obtained by considering the attenuation of the measured photoelectron intensity of the Fermi edge prior and after molecular deposition \cite{Hufner_book}. The coverage is estimated to be higher than ten layers, by considering an electron mean free path of 10 \AA{} \cite{NIST_electron_mean_free_path2010} and a nominal thickness for the SL of 3 \AA{}. This is valid under the assumption of a layer-by-layer growth and the accuracy of the value is limited by the uncertainties in the electron mean free paths in organic layers~\cite{Groburan1979}, and by the definition of the thickness for the SL \cite{massimi_JPCC_2015,Sarkar_2003_DBBA_crystal_structure}.
In the case of deposition on Au(110) kept at 470 K, the molecular SL coverage has been associated with the transition from the Au(110) 1$\times$2 to the 1$\times$4 reconstruction observed by LEED, corresponding to the opening of larger channels in the Au$[001]$ direction in order to host GNRs, as defined in Ref. \cite{massimi_JPCC_2015}. This transition was used to calibrate the temperature between the different experimental chambers. This SL coverage does not correspond to the Au(110) being completely covered by molecular species and regions of bare surface are still visible in the scanning tunneling microscopy images reported in Ref. \cite{massimi_JPCC_2015}. 

\subsection{Computational details} 
%
{\it Molecules.} Density functional theory (DFT) calculations for isolated molecules and oligomers were performed by using the ORCA package~\cite{orca}, with the aug-cc-pVTZ \cite{Dunning_89} basis set (unless differently specified). Calculations were done by using standard LDA (SVWN5)~\cite{Slater51,Vosko1980} and PBE \cite{Perdew96,Perdew97} functionals, as well as a number of hybrid and range-separated functionals (B3LYP \cite{B3LYP, pbe0}, CAM-B3LYP \cite{CAMB3LYP}, and BHandH \cite{bhh_paper}). This thorough study allows us to quantitatively compare our results with UPS data, overcoming the known limitations~\cite{chon+02jcp,onid+02rmp,ferr+14prb,nguy+15prl} of standard local and semilocal functionals (such as LDA and GGA) in estimating
spectral quantities. In addition, LDA, PBE, and PBE0 results calculated with the ORCA suite (the energy levels of isolated DBBA) were compared with those obtained by Quantum ESPRESSO~\cite{Giannozzi09}, using a plane waves basis set and pseudo-potentials. We find an overall agreement on the eigenvalues within 0.05 eV, which allows for a consistent comparison with data calculated for extended systems, as described below.

{\it Extended systems}. Periodic plane-wave calculations for polyanthryl (PA) and GNR on Au(110) were carried out using the Quantum-ESPRESSO package. The local density approximation (LDA, Perdew-Zunger parametrization\cite{perd-zung81prb}) was adopted for the exchange-correlation potential, and ultrasoft pseudo-potentials were used to model the electron-ion interaction. 
The kinetic energy cutoff for the wave functions (charge density) was set to 25 (300) Ry.
The surfaces were modelled using a 9-layers slab of Au(110) for the 1$\times$2 and the 1$\times$4 reconstructions; symmetric slabs were chosen in order to avoid unphysical dipole moments across the system. In order to accommodate two primitive cells of N7 GNR, an orthogonal $4\times 3\sqrt{2}$ 
surface cell of Au(110) was employed. 
The Brillouin zone was sampled by using a 4$\times$8$\times$1 $\mathbf{k}$-points grid both for the 1$\times$2 and the 1$\times$4 reconstruction.
A vacuum region of 12~\AA{} was added in the direction perpendicular to the slab to avoid spurious interactions with the system replicas.
The in-plane lattice parameter was set to the LDA optimized parameter for bulk Au (4.05 \AA). The atomic positions within the cell were fully optimized with a force threshold of 0.013 eV/\AA. 
With this choice for the lattice parameter, the GNR presents a residual 1\% stretching along the main axis, as compared to its free-standing optimized geometry.
Further details can be found in Ref.\cite{massimi_JPCC_2015}.
\begin{acknowledgement}
This work was funded by MIUR PRIN Grant No. 20105ZZTSE, and MAE Grant No. US14GR12. ADP acknowledges La Sapienza University for her fellowship (Grant No. 105/2014). We acknowledge partial support from the EU Centre of Excellence ``MaX'' (Grant No. 676598).
Computational resources were provided by the PRACE project on the FERMI machine at CINECA (Grant No. Pra11\_2921), and by the SOPRANO Grant by the ISCRA program (CINECA).
We kindly thank Simone Lisi and Lorenzo Massimi for experimental assistance and useful discussions.

\end{acknowledgement}
\section{Supporting Information}
Supporting Information reports ultraviolet photoemission measurements on a thin film of DBBA molecules and on DBBA molecules deposited on Au(110) kept at 470 K at increasing coverages; tests about the dependence of the electronic structure of DBBA on the localized basis set used in the calculations; a study of the length dependence of the DOS of oligo-anthryl chains; the LDA computed density of states of bianthracene adsorbed on Au(110). This material is available free of charge at http://pubs.acs.org/.

\textbf{Keywords}: Polymerization;Graphene Nanoribbons;Photoelectron spectroscopy;DBBA; Molecular orbitals 

\newpage
\begin{tocentry}
\centering
  \includegraphics[width=7.5 cm]{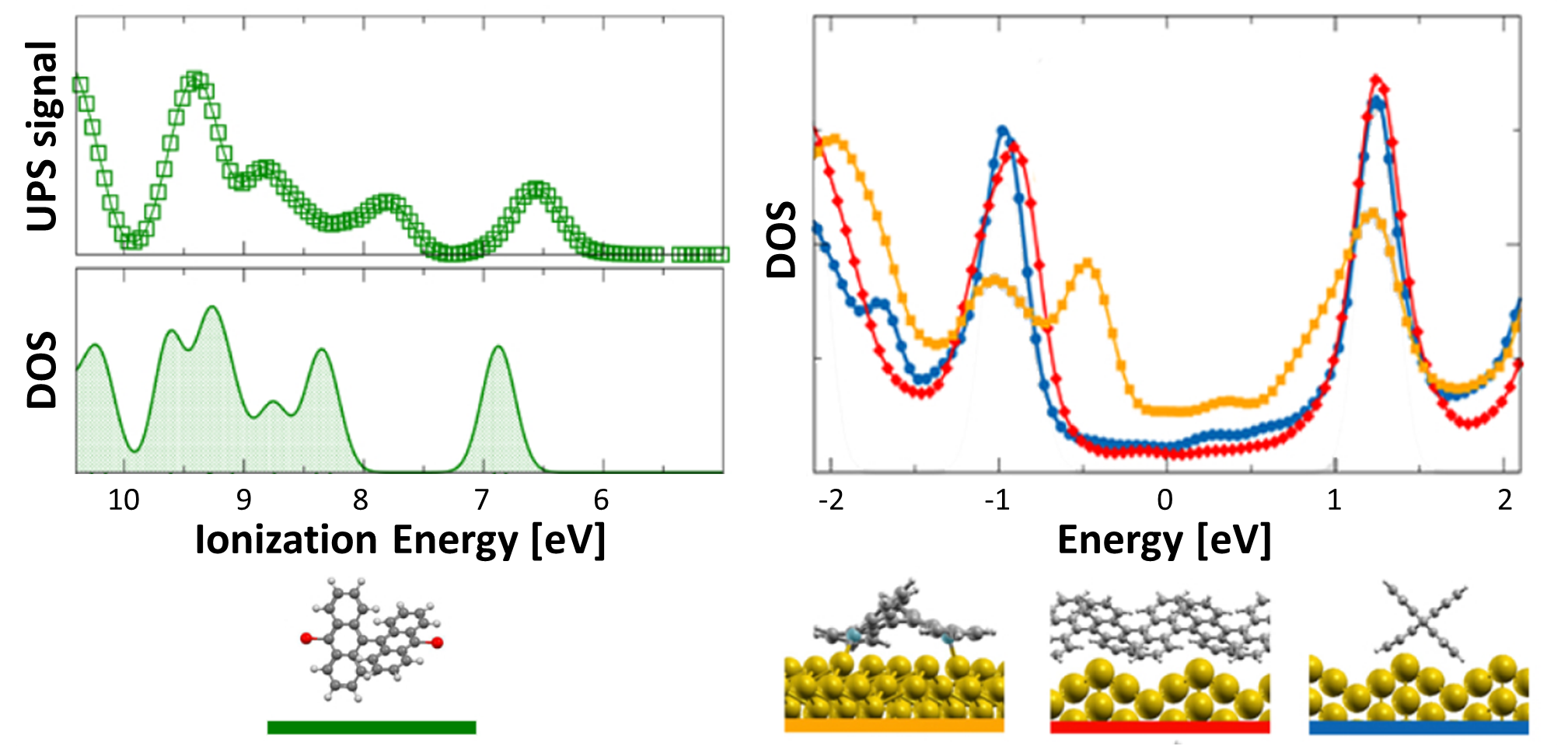}
\end{tocentry}

\bibliographystyle{achemso}

\providecommand{\latin}[1]{#1}
\providecommand*\mcitethebibliography{\thebibliography}
\csname @ifundefined\endcsname{endmcitethebibliography}
  {\let\endmcitethebibliography\endthebibliography}{}

\end{document}